\documentclass{article}
\usepackage{spconf,amssymb,amsmath,epsfig}
\usepackage{textcomp}
\usepackage{amsfonts,amsthm,amsopn}
\usepackage{bm}
\usepackage{url}
\usepackage[capitalise]{cleveref}
\usepackage{graphicx,caption,lipsum}
\usepackage{booktabs,multirow}
\usepackage{placeins}
\usepackage{algpseudocode,algorithm}
\usepackage{color}
\usepackage[table,xcdraw]{xcolor}
\usepackage{bbm}
\usepackage{enumerate}   
\usepackage{adjustbox}
\usepackage[normalem]{ulem}
\usepackage{fixltx2e}
\usepackage[toc,page]{appendix}
\usepackage{makecell}
\usepackage{soul}

\usepackage{color}
\definecolor{mypink1}{rgb}{0.858, 0.188, 0.478}

\title{Parameter-Efficient Tuning with Adapters for Speaker Verification}
\title{Efficient Adapter Tuning of Pre-trained Speech 
Models for \protect\\ Automatic Speaker Verification}

\name{{Mufan Sang, John H.L. Hansen}}
\address{
Center for Robust Speech Systems (CRSS), University of Texas at Dallas, TX, USA \\
{\small \tt \{mufan.sang, john.hansen\}@utdallas.edu}}
\newcommand{\mfmod}[1]{{\color{red}{\emph{#1}}}}
\renewcommand{\mfmod}[1]{}

\graphicspath{{./image/}}

\begin{document}

\maketitle

\ninept

\begin{abstract}
With excellent generalization ability, self-supervised speech models have shown impressive performance on various downstream speech tasks in the pre-training and fine-tuning paradigm. However, as the growing size of pre-trained models, fine-tuning becomes practically unfeasible due to heavy computation and storage overhead, as well as the risk of overfitting. Adapters are lightweight modules inserted into pre-trained models to facilitate parameter-efficient adaptation. In this paper, we propose an effective adapter framework designed for adapting self-supervised speech models to the speaker verification task. With a parallel adapter design, our proposed framework inserts two types of adapters into the pre-trained model, allowing the adaptation of latent features within intermediate Transformer layers and output embeddings from all Transformer layers. We conduct comprehensive experiments to validate the efficiency and effectiveness of the proposed framework. Experimental results on the VoxCeleb1 dataset demonstrate that the proposed adapters surpass fine-tuning and other parameter-efficient transfer learning methods, achieving superior performance while updating only 5\% of the parameters. 

% It generalization ability to various downstream speech tasks brings Transformer architecture with large amounts of unlabeled data     
\end{abstract}

%
%------------------------------------------------------------------------------------------
%------------------------------------------------------------------------------------------

%\vspace{-2.5ex}
\section{Introduction}
\vspace{-1.0ex}

In recent years, we have seen the rapid development of speaker verification (SV) driven by deep learning. Various models and methods for SV have been introduced, encompassing different deep neural network (DNN) architectures~\cite{snyder2018x, zeinali2019but, desplanques2020ecapa}, attention mechanisms~\cite{zhou2019deep}, Transformer-based architectures~\cite{zhang22h_interspeech, sang2023improving}, and self-supervised SV systems~\cite{chen2022unispeech, sang2022self, zhang2021contrastive}. Most of these works focus on utilizing task-specific datasets to train SV systems from scratch. \mfmod{Recently, the emergence of large-scaled pre-trained models has propelled the research in the field of natural language processing (NLP) \cite{kenton2019bert,wang2023dual}}Recently, the emergence of large-scaled pre-trained speech models has propelled the research in the field of speech processing. Taking the advantages of Transformer architecture, self-supervised learning (SSL), and increasingly large amounts of unlabeled data, pre-trained models exhibit strong generalization capabilities across various downstream speech tasks. Applying large-scale pre-trained speech models (e.g., HuBERT~\cite{hsu2021hubert}, WavLM~\cite{chen2022wavlm}) to downstream tasks has remarkably improved performance over conventional models. The question of how to more efficiently utilize pre-trained models to improve the performance of downstream tasks remains an open area for investigation. 

The pre-training and fine-tuning paradigm has become the most common approach for adapting pre-trained models to downstream tasks~\cite{chen2022wavlm, kenton2019bert,wang2023dual}. However, fine-tuning has drawn some issues due to two main reasons. Firstly, fine-tuning requires one to update all the model parameters, store and deploy a separate copy of the model parameters for each individual downstream task. As the size of pre-trained SSL models increases, fine-tuning becomes prohibitively costly in terms of training, storage, and deployment, rendering it practically infeasible.\mfmod{to achieve competitive performance for different downstream tasks, pre-trained models need to be retrained and the majority of all parameters of pre-trained models need to be updated. After fine-tuning, the parameters of the whole pre-trained models need to be stored.}\mfmod{However, this strategy requires one to store and deploy a separate copy of the model parameters for every single task. This is an expensive and often infeasible proposition, especially for modern Transformer- based architectures, which are significantly larger than their convolutional neural networks (ConvNet) counterparts}\mfmod{Another go-to strategy is performing linear-probing by stacking an additional trainable multi-layer perceptron (MLP) layer in the end. It is parameter-efficient yet suboptimal in terms of performance. In fact, fine-tuning all the parameters of a large-scale pre-trained model and storing separate instances for different downstream tasks are practically infeasible. With the increasing size of pre-trained SSL models, fine-tuning brings prohibitive adaptation costs for training, storage, and deployment costs become incredibly high for downstream fine-tuning.} Secondly, pre-trained models are prone to overfitting when fine-tuned on limited amounts of data for downstream tasks, which degrades their generalization abilities. Therefore, parameter-efficient fine-tuning methods are crucial for large-scale pre-trained model adaptation. \mfmod{A more challenging yet less studied problem is whether one can achieve better performance than fine-tuning with fewer parameters. }A simple and straightforward approach is linear probing, where the pre-trained model remains fixed and the stacked classification head is fine-tuned for each downstream task. \mfmod{As one of the transfer learning methods, linear probing is simple and straightforward by stacking and tuning an additional trainable linear layer and freezing the whole pre-trained model.}However, it often results in unsatisfactory performance compared to full fine-tuning.\mfmod{It requires updating much fewer parameters yet is suboptimal in terms of performance. } More recently, adapters have drawn more and more attention for transferring knowledge from pre-trained models to downstream tasks. Adapters~\cite{houlsby2019parameter} were first proposed in the Natural Language Processing (NLP) field for model adaptation, which inserts lightweight modules with bottleneck architecture into Transformer layers after multi-head self-attention (MHSA) and feed-forward network (FFN) modules. A bottleneck layer consists of a down and up projection pair that shrinks and recovers the size of token hidden states. During fine-tuning, only the inserted adapters get updated and other parts of the model keep frozen. Some studies~\cite{thomas2022efficient,fan2022draft,chen2023chapter} explored the use of adapters to adapt pre-trained models to diverse speech processing tasks. However, most of them do not adequately utilize the information embedded in different layers of pre-trained models. Efficient methods for adapting pre-trained models to speaker verification are not well-studied.       

\mfmod{Adapter can be a As one of the most common approaches, adapter. To this end, a branch of parameter-efficient methods for model tuning arises. Generally, delta tuning only updates a small number of parameters (inherently in the model or additionally introduced) while freezing the remaining parameters that account for the vast majority. Adapter tuning (Houlsby et al., 2019) is among the earliest approaches to steer pre-trained models with a limited number of parameters. It inserts lightweight modules with bottleneck architecture after multi-head attention and feed-forward network (FFN) layers in PLMs and only these inserted modules get updated during fine-tuning}

\mfmod{replaces the original MLP block with two branches, including the frozen and the trainable branch in parallel for adapting vision transformers to a large video action recognition, which avoids catastrophic interference with each other}

%\vspace{-4.0ex}
In this paper, we propose an effective adapter framework that consists of two modules: the Inner-layer Adapter and the Inter-layer Adapter, aiming to efficiently transfer the universal knowledge of pre-trained SSL model to the speaker verification task. The proposed adapters learn task-specific knowledge for speaker verification by adapting latent features within intermediate Transformer layers and output embeddings from all Transformer layers of pre-trained model. Moreover, we introduce a parallel adapter design that inserts and sets adapters in parallel to the FFN of Transformer layers. A scaling operation is introduced to control adapter outputs, and balance task-agnostic and task-specific features learned from original FFN branches and adapter branches within Transformer blocks. Experimental results demonstrate that our proposed adapter-tuning method significantly outperforms other transfer learning methods and full fine-tuning while updating only 5.0\% extra parameters. The primary contributions of this paper can be summarized as follows: (1) We propose an effective adapter framework that fully leverages speaker-related information embedded in different layers of pre-trained model. (2) We propose a parallel adapter design that helps the pre-trained model learn complementary task-specific knowledge. (3) We conduct comprehensive experiments to validate the efficiency and effectiveness of the proposed adapter framework.  

\vspace{-1.0ex}
\begin{figure*}[th]
\centering
\scalebox{0.99}
{
\includegraphics[width=17.8cm,height=5.8cm]{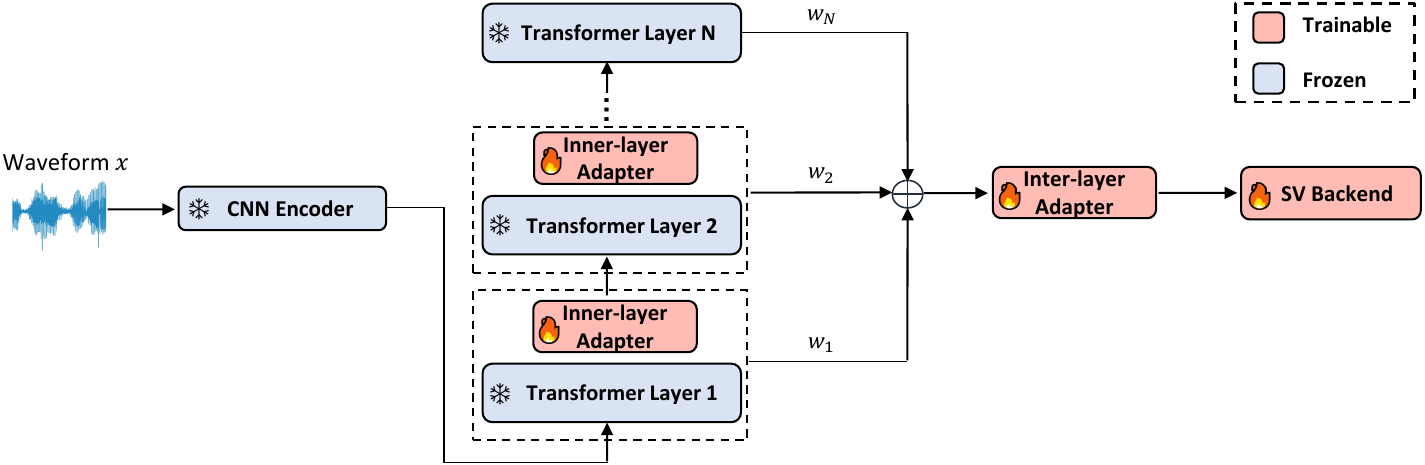}
}
%\vspace{-2.0mm}
\caption{Overview of the pre-trained model and the proposed adapter framework. During fine-tuning, the pre-trained model is frozen, only the Inner-layer Adapter, Inter-layer Adapter, and the SV backend are updated.} 
\label{fig:system}
\end{figure*}

%\vspace{-1.0ex}
\section{Related Works}
%\vspace{-1.0ex}
\subsection{Self-supervised Pre-trained Speech Models}
\mfmod{CNNs and RNNs has been used for a long time as the main module to extract features in audio and speech tasks. Recently, Transformers [] was proposed and have shown remarkable performance in various fields on natural language processing and computer vision tasks. For Transformers, the self-attention mechanisms capture the global contextual relationship between inputs. In speech field, Transformers}

\mfmod{The remarkable success of large-scale transformer models in NLP has sparked a growing interest in adopting these models in speech field. [] proposed Wav2vec with a discriminative learning objective. It uses the contrastive InfoNCE loss to discriminate the correlated positive samples from negative samples. Following the direction, [] proposed Wav2vec 2.0 with feature masking to identify the true quantized latent speech representation for a masked time step within a set of distractors. [] proposed HuBERT with a predict discrete targets of masked region uses an offline clustering step to generate noisy labels for Masked Language Model pretraining. }

Self-supervised learning (SSL) can utilize large amounts of unlabeled data to help models learn generic representations, thereby it has received increasing attention in the speech field. Recently, various SSL based pre-trained models and methods, including wav2vec~\cite{schneider2019wav2vec}, wav2vec 2.0~\cite{baevski2020wav2vec}, HuBERT~\cite{hsu2021hubert}, and WavLM~\cite{chen2022wavlm} have been proven effective on certain speech tasks \cite{thomas2022efficient,chen2023chapter,shan2023phoneme}. Among them, WavLM was proposed to explore a full stack of speech tasks instead of focusing on specific tasks. It combines masked speech prediction and denoising during pre-training to learn not only knowledge related to automatic speech recognition (ASR) but also information about other non-ASR tasks. \mfmod{In summary, these SSL pre-trained models have remarkably advanced the development of various speech fields \cite{thomas2022efficient,chen2023chapter,shan2023phoneme} with excellent generalization ability. }The pre-train and fine-tune paradigm has shown its success in adapting pre-trained models to downstream speech tasks. However, fine-tuning large-scale pre-trained models remains data-dependent and computationally expensive, limiting the broader application of SSL pre-trained models. Therefore, it is worthwhile to explore how to transfer the knowledge of pre-trained models to downstream tasks with lower computation and storage costs. 

\subsection{Adapter-based Tuning}
Adapters~\cite{houlsby2019parameter} were initially introduced as an alternative approach for adapting large-scale pre-trained language models in NLP. Adapters modify the feature extractors by inserting some lightweight bottleneck modules without changing the parameters of pre-trained models. Adapter-based methods have proven to be comparable with full ﬁne-tuning with much higher parameter efficiency, and sometimes perform slightly better in low-resource settings~\cite{he2022towards}. With the advantage, adapters have also been applied to computer vision tasks~\cite{sung2022vl,chen2022adaptformer}.  

\mfmod{In the field of speech processing, }Recently, adapters have also been introduced in speech processing tasks. In~\cite{kannan2019large}, researchers applied adapters to the RNN-T model for multilingual ASR. The work~\cite{winata2020adapt} proposed employing adapters to a speech Transformer to address the long-tail problem of multilingual ASR. In~\cite{thomas2022efficient}, adapters were applied to wav2vec 2.0 to increase the model's scalability to multiple languages. In~\cite{fan2022draft}, adapters were utilized to improve the domain adaptation of SSL models, including wav2vec 2.0 and HuBERT for child ASR. SimAdapter~\cite{hou2021exploiting} was proposed for cross-lingual low-resource ASR. In~\cite{chen2023chapter,otake2023parameter,peng2023parameter}, researchers explored the effectiveness of adapters for different downstream speech tasks beyond ASR (e.g., emotion recognition, speaker verification, intent classiﬁcation). However, most of these studies do not sufficiently leverage the information embedded in different layers of pre-trained models. Thus, the goal of this study is to design an efficient and effective adapter framework for speaker verification task.

%\vspace{-2.0ex}
\section{Method}
%\section{ADAPTER ARCHITECTURE FOR SPEAKER RECOGNITION}
%\vspace{-1.0ex}
We propose a novel adapter framework to efﬁciently transfer the knowledge of large pre-trained speech models to speaker verification task. In our framework, we insert two types of adapters into the pre-trained backbone model: (1) Inner-layer Adapter, inserted within the intermediate Transformer layers. (2) Inter-layer Adapter, inserted after the weighed-sum operation between the backbone and speaker verification backend. \mfmod{inserted the Inner-layer adapters are inserted within the intermediate Transformer layers and the Inter-layer Adapter is inserted after the weighed-sum operation between the pre-trained backbone and speaker verification backend. }The overall framework is illustrated in Fig. 1.   
    
\subsection{Inner-layer Adapter and Inter-layer Adapter}
%\vspace{-1.0ex}
% Pre-trained self-supervised models utilize massive amounts of unlabeled speech data to learn rich representations of spoken languages. Thus, they can be utilized as the backbone with different downstream heads for various downstream speech tasks. Fine-tuning requires updating all parameters of pre-trained models, which causes high computational cost. As one of the parameter-efficient transfer learning methods, adapter-tuning inserts lightweight modules (called adapters)  

Adapters are lightweight modules inserted into the Transformer layers of pre-trained models for adaptation. To preserve the generalization ability of pre-trained models, only adapters are fine-tuned, and the pre-trained model keeps frozen during training.\mfmod{The study in~\cite{yang2021superb} indicates that not only the last layer but also other lower layers can contribute to better performance for pre-trained models on various downstream speech tasks. } To better utilize output representations from all intermediate layers, we propose the Inner-layer Adapter and Inter-layer Adapter, allowing the adaptation of latent features within intermediate Transformer layers and output embeddings from all Transformer layers.\mfmod{facilitate adapt latent representation within Transformer layers and weighed sum of output embeddings across all Transformer layers. }

In many studies~\cite{houlsby2019parameter,thomas2022efficient}, adapters are inserted after both multi-head self-attention and feed-forward network. To improve parameter efficiency, we insert the Inner-layer Adapter after FFN only. The Inner-layer Adapter has a bottleneck structure consisting of a down-projection to reduce the hidden dimension $d$ to bottleneck dimension $\hat{d}$ with parameter $\boldsymbol{W}_{\text {down}}\in R^{d \times \hat{d}}$, an up-projection with parameter $\boldsymbol{W}_{\text {up}}\in R^{\hat{d} \times d}$, a non-linear activation function between them, layer normalization (LN), and a residual connection. Given $\boldsymbol{x_{i}}$ as the input feature of FFN, the output of Inner-layer Adapter can be formulated as:
\begin{equation}
\begin{aligned}
\tilde{\boldsymbol{z}^s_i} &=\mathrm{FFN}\left(\boldsymbol{x_i}\right)+\mathrm{LN}(\boldsymbol{W}_{\text {up}}f\left(\boldsymbol{W}_{\text {down }}\mathrm{FFN}\left(\boldsymbol{x_i}\right)\right))
\end{aligned}
\end{equation}
where $f$ denotes the ReLU activation function. 

The previous study~\cite{yang21c_interspeech} indicates that the output representations from lower layers of pre-trained models can contribute to better performance on various downstream speech tasks. Consequently, we add a group of trainable weights to average the output representations from all layers. The Inner-layer Adapters are integrated into intermediate Transformer layers for adapting latent features within layers explicitly. However, the interaction among all layers is ignored. To better adapt the pre-trained model and fully leverage the speaker-related information embedded in all layers, we propose the Inter-layer Adapter. As shown in Fig. 1, we insert the Inter-layer Adapter after the weighted sum operation to facilitate the model adaptation. The Inter-layer Adapter consists of a fully connected (FC) layer and a non-linear activation function with LN. Given the output representation from the $i$-th layer as $\boldsymbol H_{i}$, the output of the Inter-layer Adapter is computed as:
\begin{equation}
\begin{aligned}
\boldsymbol{\tilde{H}}=\mathrm{LN}(f(\boldsymbol{W}_{\text {inter}}(\sum_{i=1}^{N} w_{i} \boldsymbol H_{i}\label{con:DCT1D2})))
\end{aligned}
\end{equation}
where $\boldsymbol{W}_{\text {inter}}\in R^{d \times e}$ denotes the FC layer of Inter-layer Adapter, $d$ is the hidden dimension and $e$ is the speaker embedding dimension, $f$ denotes the ReLU activation function and $w_{i}$ denotes the trainable weight for the $i$-th layer. \mfmod{In this way, the proposed framework}

\subsection{Parallel Adapter Design}
%\vspace{-1.0ex}
Adapters are usually inserted sequentially after MHSA and FFN, and take their outputs as inputs for further computing. Inspired by~\cite{chen2022adaptformer,he2022towards}, we propose a parallel design for our adapters and illustrate it in Fig. 2. Unlike sequential design, the parallel adapter is integrated into an additional sub-branch for task-specific fine-tuning. The output of the parallel adapter is rescaled by a factor $s$ and then added to the original branch through a residual connection. The scaling factor $s$ is proposed to control the balance between the task-agnostic features obtained from the original frozen branch and the task-specific features obtained from the tunable adapter branch. This parallel design enables the pre-trained model to preserve its generalization capability, while the domain-specific features learned from the adapters can serve as a valuable complement for feature ensemble.\mfmod{In this way, the pre-trained model can keep its generalization ability and the domain-specific features learned from adapters can be a supplementation for feature ensemble.} For a specific input feature of FFN $\boldsymbol{x_{i}}$, the output of the parallel adapter is formulated as: 
\begin{equation}
\begin{aligned}
\tilde{\boldsymbol{z}^{p}_i} &=\mathrm{LN}(\boldsymbol{W}_{\text {up}}f\left(\boldsymbol{W}_{\text {down }}\boldsymbol{x_i}\right))
\end{aligned}
\end{equation}
Accordingly, features from the adapter branch, FFN branch and the residual connection are fused, and the final output of the $i$-th Transformer layer is shown as:
\begin{equation}
\begin{aligned}
{\boldsymbol H_{i}} &=\mathrm{LN}\left(\mathrm{FFN}\left(\boldsymbol{x_i}\right)+s \cdot \tilde{\boldsymbol{z}^{p}_i}+\boldsymbol{x_i}\right)
\end{aligned}
\end{equation}
% final output before the last LN is written as:

\begin{figure}[t]
\centering
\scalebox{1.0}
{
\includegraphics[width=8.5cm,height=6.0cm]{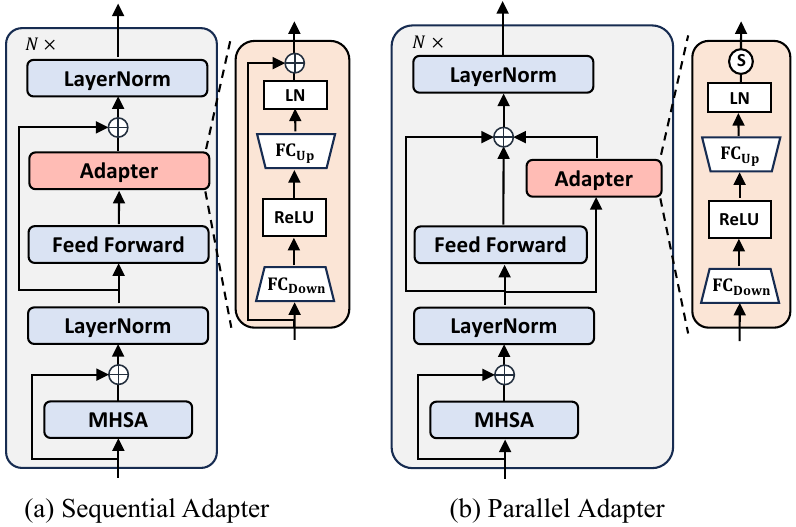}
}
%\vspace{-6.0mm}
\caption{Detailed architectures of (a) Sequential Adapter and (b) Parallel Adapter.} 
\label{fig:Swin}
\end{figure} 

%\vspace{-2.0ex}
\section{Experiments}
%\vspace{-1.5ex}
\subsection{Datasets}
%\subsubsection{VoxCeleb}
%\vspace{-1.0ex}
The speaker verification systems are trained and evaluated on the VoxCeleb1~\cite{nagrani2017voxceleb} dataset, which contains 148,642 utterances from 1,211 speakers in the development set and 4,874 utterances from 40 speakers in the test set. We report the performance of systems on the VoxCeleb1-O evaluation trial. 

Moreover, we evaluate the proposed adapter framework in more challenging scenarios on the naturalistic $1^{\rm{st}}$48-UTD forensic corpus~\cite{sang2020open}, where the total duration is 3.5 hours and more than 50\% of utterances are shorter than 2 seconds. The training set consists of 3,755 utterances from 228 speakers, and the test set contains 882 utterances from 39 speakers. More details of the corpus can be found in~\cite{sang2020open}.
% \subsection{Compared Methods}

%\vspace{-2.0ex}
\subsection{Implementation Details}
\mfmod{Attentive statistic pooling  (ASP)~\cite{okabe2018attentive} is used to generate utterance-level embeddings. The models are trained with additive margin softmax (AM-softmax) loss~\cite{wang2018cosface} with a margin of 0.2 and a scaling factor of 30.}
In this study, we employ the pre-trained WavLM Base+ as the backbone model. It comprises a convolutional feature encoder and 12 Transformer blocks equipped with gated relative position bias. \mfmod{The convolutional feature encoder comprises seven blocks of temporal convolution followed by layer normalization and a GELU activation layer []. }Within the Transformer blocks, there are 8 attention heads, each with 768-dimensional hidden states. The WavLM Base+ has 94.70 million parameters. The speaker verification backend is composed of an average time pooling layer and two FC layers, with an embedding size of 512. The Inner-layer Adapter consists of two FC layers with a bottleneck dimension of 256, a ReLU activation function between them, LN and a residual connection. The Inter-layer Adapter consists of a FC layer with a hidden dimension of 512, followed by a ReLU activation function and LN.     

The models are trained with the cross-entropy loss. We use the Adam~\cite{kingma2014adam} optimizer with an initial learning rate of 5e-4 for SV backend and 1e-5 for all other parameters. We apply a warm-up strategy at the first 38k steps and the learning rates decrease to 2.5e-5 for SV backend and 5e-7 for all other parameters in the remaining steps.

We compare our method with several transfer learning approaches, including full fine-tuning, linear probing, weighted sum, and two adapter-based methods: Houlsby adapter~\cite{houlsby2019parameter} and E-adapter and L-adapter (E+L adapter)~\cite{otake2023parameter}. In this study, in the case of full fine-tuning, we update all the parameters of WavLM Base+ but keep its convolutional encoder frozen. The weighted sum method is implemented as in~\cite{yang21c_interspeech}, which has been demonstrated to be effective for speaker verification. To ensure a fair comparison, we reimplement the Houlsby adapter and the E+L adapter, and apply the same training configurations for all these methods.     

We report the system performance using two evaluation metrics: Equal Error Rate (EER) and minimum Detection Cost Function (minDCF) with $p_{target}=0.05$.

\begin{table}[t]
\caption{Performance of our method on VoxCeleb1. The second column represents the number of trainable parameters in the pre-trained model. Upper block: different tuning methods; Lower block: our proposed method and variants; FT: full fine-tuning; LP: linear probing.}
\setlength{\tabcolsep}{1.8mm}{
\renewcommand\arraystretch{1.3}
\scalebox{0.99}{
\begin{tabular}{lccc}
\hline \multirow{2}{*}{\textbf{Method} } & \multirow{2}{*}{ \textbf{\# Params} } & \multicolumn{2}{c}{ \textbf{VoxCeleb1-O} } \\
\cline { 3 - 4 } & & \textbf{EER $(\%)$} & \textbf{minDCF} \\
\hline FT & $85.1 \mathrm{M}$ $(90.0\%)$ & 3.69 & 0.265 \\
LP & $0.0 \mathrm{M}$ $(0.0 \%)$ & 8.58 & 0.622 \\
Weighted Sum & $0.03 \mathrm{M} $ $(0.03 \%)$ & 4.81 & 0.324 \\
Houlsby Adapter {~\cite{houlsby2019parameter}} & $9.5 \mathrm{M}$ $(10.1 \%)$ & 3.67 & 0.244 \\
E+L Adapter {~\cite{otake2023parameter}} & $9.1 \mathrm{M}$ $(9.6 \%)$ & 2.77 & 0.195 \\
\hline Ours (Inner-layer) & $4.4 \mathrm{M}$ $(4.6 \%)$ & 3.24 & 0.242 \\
Ours (Inter-layer) & $0.4 \mathrm{M}$ $(0.4 \%)$ & 3.01 & 0.215 \\
Ours (Inner+Inter) & $4.8 \mathrm{M}$ $(5.0 \%)$ & $\mathbf{2.58}$ & $\mathbf{0.187}$ \\
\hline
\end{tabular}}
}
\end{table}

% might cause optimization difﬁculty.
\subsection{Comparison among Transfer Learning Methods}
%\subsection{Evaluation on VoxCeleb Dataset}
In the experiments, we investigate the performance of our proposed adapter framework and evaluate it on the VoxCeleb1 dataset. We compare our method with other transfer learning methods, including fine-tuning, linear-probing, weighted sum, and two adapter-based methods: Houlsby adapter and E+L adapter. From Table 1, we can observe that the proposed Inner+Inter Adapter achieves the best performance, and outperforms fine-tuning and all other methods. Notably, compared to fine-tuning, our method improves the performance with relative 30.1\% and 29.4\% reductions in EER and minDCF respectively, by introducing only 5.0\% of the pre-trained model parameters. The weighted sum fails to attain similar performance as fine-tuning, and linear probing performs significantly worse than fine-tuning. Compared to the other two adapter-based methods, our approach remarkably outperforms Houlsby adapter and E+L adapter while saving approximately 50\% of the parameters, which sufficiently demonstrates the effectiveness and parameter efficiency of the proposed adapter framework.  

Furthermore, in the lower section of Table 1, we can observe that the two variants, which insert the Inner-layer Adapter and Inter-layer Adapter separately, outperform both fine-tuning and Houlsby adapter. Compared to the weighted sum method, the Inter-layer Adapter inserts a single adapter after the weighted sum operation, achieving significantly better performance with a 37.4\% reduction in EER on VoxCeleb1-O. Specifically, our Inter-layer Adapter surpasses even the Houlsby adapter and the other variant (Inner-layer) in terms of performance while saving 22$\times$ and 10$\times$ parameters, respectively. Consequently, experimental results indicate that the Inter-layer Adapter can serve as an essential module for adapter-based methods in speaker verification.     

% \vspace{-2.5ex}
% \subsubsection{Evaluation Metrics}
% \vspace{-1.0ex}
\mfmod{Cosine similarity is adopted for scoring in the testing phase.}

\begin{table}[t]
\caption{Performance of sequential adapter and parallel adapter with learnable and fixed scaling factors.}
\vspace{-4.0mm}
\setlength{\tabcolsep}{4mm}{
\begin{center}
\renewcommand\arraystretch{1.05}

\begin{tabular}{l|cc}
\hline \makecell{\textbf{Scales}} & \textbf{EER} (\%) & \textbf{minDCF} \\
\hline \makecell[c]{Sequential} & $3.05$ & $0.208$ \\ 
\makecell{Learnable} & $2.76$ & $0.190$ \\
\makecell{0.05} & $3.45$ & $0.236$ \\
\makecell{0.1} & $3.32$ & $0.224$ \\
\makecell{0.5} & $\mathbf{2.58}$ & $\mathbf{0.187}$ \\
\makecell{1.0} & $2.69$ & $0.195$ \\
\makecell{1.5} & $2.79$ & $0.194$ \\
\makecell{2.0} & $2.82$ & $0.202$ \\
% \makecell{1.2} & $2.98$ & $0.205$ \\
% \makecell{1.4} & $\mathbf{2.64}$ & $\mathbf{0.168}$ \\
% \makecell{1.6} & $2.64$ & $0.168$ \\

\hline
\end{tabular}
\end{center}}
%\vspace{-3mm}
\label{table:KingASR}
\end{table}

%\vspace{-4.0ex}
\begin{table}[t]
\caption{Performance of different transfer learning methods on $1^{st}$48-UTD forensic dataset.}
\vspace{-4.0mm}
\setlength{\tabcolsep}{4mm}{
\begin{center}
\renewcommand\arraystretch{1.0}

\begin{tabular}{l|cc}
\hline \makecell{\textbf{Systems}} & \textbf{EER} (\%) & \textbf{minDCF} \\
\hline \makecell{FT} & $19.13$ & $0.842$ \\
\makecell{LP} & $18.52$ & $0.829$ \\
\makecell{Weighted Sum} & $18.10$ & $0.819$ \\
\makecell{Ours (Inner+Inter)} & $\mathbf{16.75}$ & $\mathbf{0.734}$ \\

\hline
\end{tabular}
\end{center}}

%\vspace{-3mm}
\label{table:KingASR}
\end{table}
%\vspace{-1.0ex}
 
\subsection{Ablation Study}
We further conduct experiments to study the effectiveness of the parallel design. We compare the performance of our adapter framework using sequential and parallel insertion formulation. As presented in Table 2, the parallel adapter yields better performance than the sequential counterpart when using learnable and fixed scaling factors ($s \geq 0.5$). Additionally, we explicitly study the impact of scaling factor on the parallel adapter. In Table 2, we observe that our parallel adapter achieves the best performance with the fixed scale at 0.5, and using a learnable scale factor results in a slightly worse but comparable performance. Increasing or decreasing the value of $s$ brings a performance drop. The reason could be that a smaller $s$ might diminish the impact of task-specific features learned from adapters, and a larger $s$ might weaken the contribution of task-agnostic features learned from the frozen pre-trained backbone. Based on the experimental results, we infer that the parallel adapter proves to be a more suitable choice for speaker verification. 
\mfmod{it can be concluded that the parallel adapter can be a better choice for speaker verification.}     

% The parallel adapter is inserted in an independent branch instead of added sequentially after the FFN. In this way, the pre-trained model keeps the original branch            

%\vspace{-2mm}

%\vspace{-1.0ex}
\subsection{Evaluation in More Challenging Scenarios}
In this section, we evaluate the performance of the proposed method on a more challenging dataset for forensic speaker recognition. As shown in Table 3, our method consistently outperforms fine-tuning, linear probing, and weighted sum, which illustrates the effectiveness and robustness of the proposed method even in more complex scenarios.   
%\vspace{-1.0ex}

%\vspace{-1.0ex}
\section{Conclusions}
\vspace{-0.5ex}
In this paper, we propose a parameter-efficient adapter-tuning framework aimed at effectively transferring the knowledge of pre-trained self-supervised speech models to speaker verification task. To sufficiently leverage the information embedded in all intermediate layers, our framework incorporates Inner-layer Adapters after the feed-forward network to adapt latent features within Transformer blocks. Additionally, it inserts an Inter-layer Adapter after the weighted sum operation to adapt the aggregated hidden representations extracted from all layers. The parallel design further improves model performance. Experimental results show that the proposed adapter outperforms fine-tuning and other transfer learning methods while updating only 5\% extra parameters. The proposed framework can efficiently adapt the pre-trained model to the speaker verification task, leading to substantial reductions in computational and storage costs. We hope this work will inspire future research on parameter-efficient transfer learning of large-scale pre-trained speech models for speaker verification.  
% without large-scale pre-training. 

\vfill\pagebreak

% \newpage
\footnotesize
%\small
\bibliographystyle{IEEEbib}
\bibliography{icassp21}

\begin{thebibliography}{10}

\bibitem{snyder2018x}
David Snyder, Daniel Garcia-Romero, Gregory Sell, Daniel Povey, and Sanjeev Khudanpur,
\newblock ``X-vectors: Robust dnn embeddings for speaker recognition,''
\newblock in {\em ICASSP}, 2018, pp. 5329--5333.

\bibitem{zeinali2019but}
Hossein Zeinali, Shuai Wang, Anna Silnova, Pavel Mat{\v{e}}jka, and Old{\v{r}}ich Plchot,
\newblock ``But system description to voxceleb speaker recognition challenge 2019,''
\newblock {\em arXiv preprint arXiv:1910.12592}, 2019.

\bibitem{desplanques2020ecapa}
Brecht Desplanques, Jenthe Thienpondt, and Kris Demuynck,
\newblock ``Ecapa-tdnn: Emphasized channel attention, propagation and aggregation in tdnn based speaker verification,''
\newblock {\em arXiv preprint arXiv:2005.07143}, 2020.

\bibitem{zhou2019deep}
Jianfeng Zhou, Tao Jiang, Zheng Li, Lin Li, and Qingyang Hong,
\newblock ``Deep speaker embedding extraction with channel-wise feature responses and additive supervision softmax loss function.,''
\newblock in {\em Interspeech}, 2019, pp. 2883--2887.

\bibitem{zhang22h_interspeech}
Yang Zhang, Zhiqiang Lv, Haibin Wu, Shanshan Zhang, Pengfei Hu, Zhiyong Wu, Hung yi~Lee, and Helen Meng,
\newblock ``{MFA-Conformer: Multi-scale Feature Aggregation Conformer for Automatic Speaker Verification},''
\newblock in {\em Proc. Interspeech 2022}, 2022, pp. 306--310.

\bibitem{sang2023improving}
Mufan Sang, Yong Zhao, Gang Liu, John~HL Hansen, and Jian Wu,
\newblock ``Improving transformer-based networks with locality for automatic speaker verification,''
\newblock in {\em ICASSP 2023-2023 IEEE International Conference on Acoustics, Speech and Signal Processing (ICASSP)}. IEEE, 2023, pp. 1--5.

\bibitem{chen2022unispeech}
Sanyuan Chen, Yu~Wu, Chengyi Wang, Zhengyang Chen, Zhuo Chen, Shujie Liu, Jian Wu, Yao Qian, Furu Wei, Jinyu Li, et~al.,
\newblock ``Unispeech-sat: Universal speech representation learning with speaker aware pre-training,''
\newblock in {\em ICASSP 2022-2022 IEEE International Conference on Acoustics, Speech and Signal Processing (ICASSP)}. IEEE, 2022, pp. 6152--6156.

\bibitem{sang2022self}
Mufan Sang, Haoqi Li, Fang Liu, Andrew~O Arnold, and Li~Wan,
\newblock ``Self-supervised speaker verification with simple siamese network and self-supervised regularization,''
\newblock in {\em ICASSP}. IEEE, 2022, pp. 6127--6131.

\bibitem{zhang2021contrastive}
Haoran Zhang, Yuexian Zou, and Helin Wang,
\newblock ``Contrastive self-supervised learning for text-independent speaker verification,''
\newblock in {\em ICASSP}. IEEE, 2021, pp. 6713--6717.

\bibitem{hsu2021hubert}
Wei-Ning Hsu, Benjamin Bolte, Yao-Hung~Hubert Tsai, Kushal Lakhotia, Ruslan Salakhutdinov, and Abdelrahman Mohamed,
\newblock ``Hubert: Self-supervised speech representation learning by masked prediction of hidden units,''
\newblock {\em IEEE/ACM Transactions on Audio, Speech, and Language Processing}, vol. 29, pp. 3451--3460, 2021.

\bibitem{chen2022wavlm}
Sanyuan Chen, Chengyi Wang, Zhengyang Chen, Yu~Wu, Shujie Liu, Zhuo Chen, Jinyu Li, Naoyuki Kanda, Takuya Yoshioka, Xiong Xiao, et~al.,
\newblock ``Wavlm: Large-scale self-supervised pre-training for full stack speech processing,''
\newblock {\em IEEE Journal of Selected Topics in Signal Processing}, vol. 16, no. 6, pp. 1505--1518, 2022.

\bibitem{kenton2019bert}
Jacob Devlin Ming-Wei~Chang Kenton and Lee~Kristina Toutanova,
\newblock ``Bert: Pre-training of deep bidirectional transformers for language understanding,''
\newblock in {\em Proceedings of NAACL}, 2019, pp. 4171--4186.

\bibitem{wang2023dual}
Yigong Wang, Zhuoyi Wang, Yu~Lin, Jinghui Guo, Sadaf Halim, and Latifur Khan,
\newblock ``Dual contrastive learning framework for incremental text classification,''
\newblock in {\em Findings of the Association for Computational Linguistics: EMNLP 2023}, 2023, pp. 194--206.

\bibitem{houlsby2019parameter}
Neil Houlsby, Andrei Giurgiu, Stanislaw Jastrzebski, Bruna Morrone, Quentin De~Laroussilhe, Andrea Gesmundo, Mona Attariyan, and Sylvain Gelly,
\newblock ``Parameter-efficient transfer learning for nlp,''
\newblock in {\em International Conference on Machine Learning}. PMLR, 2019, pp. 2790--2799.

\bibitem{thomas2022efficient}
Bethan Thomas, Samuel Kessler, and Salah Karout,
\newblock ``Efficient adapter transfer of self-supervised speech models for automatic speech recognition,''
\newblock in {\em ICASSP 2022-2022 IEEE International Conference on Acoustics, Speech and Signal Processing (ICASSP)}. IEEE, 2022, pp. 7102--7106.

\bibitem{fan2022draft}
Ruchao Fan and Abeer Alwan,
\newblock ``Draft: A novel framework to reduce domain shifting in self-supervised learning and its application to children's asr,''
\newblock {\em arXiv preprint arXiv:2206.07931}, 2022.

\bibitem{chen2023chapter}
Zih-Ching Chen, Yu-Shun Sung, and Hung-yi Lee,
\newblock ``Chapter: Exploiting convolutional neural network adapters for self-supervised speech models,''
\newblock in {\em 2023 IEEE International Conference on Acoustics, Speech, and Signal Processing Workshops (ICASSPW)}. IEEE, 2023, pp. 1--5.

\bibitem{schneider2019wav2vec}
Steffen Schneider, Alexei Baevski, Ronan Collobert, and Michael Auli,
\newblock ``wav2vec: Unsupervised pre-training for speech recognition,''
\newblock {\em Proc. Interspeech 2019}, pp. 3465--3469, 2019.

\bibitem{baevski2020wav2vec}
Alexei Baevski, Yuhao Zhou, Abdelrahman Mohamed, and Michael Auli,
\newblock ``wav2vec 2.0: A framework for self-supervised learning of speech representations,''
\newblock {\em Advances in neural information processing systems}, vol. 33, pp. 12449--12460, 2020.

\bibitem{shan2023phoneme}
Siyuan Shan, Yang Li, Amartya Banerjee, and Junier~B Oliva,
\newblock ``Phoneme hallucinator: One-shot voice conversion via set expansion,''
\newblock {\em arXiv preprint arXiv:2308.06382}, 2023.

\bibitem{he2022towards}
Junxian He, Chunting Zhou, Xuezhe Ma, Taylor Berg-Kirkpatrick, and Graham Neubig,
\newblock ``Towards a unified view of parameter-efficient transfer learning,''
\newblock {\em International Conference on Learning Representations}, 2022.

\bibitem{sung2022vl}
Yi-Lin Sung, Jaemin Cho, and Mohit Bansal,
\newblock ``Vl-adapter: Parameter-efficient transfer learning for vision-and-language tasks,''
\newblock in {\em Proceedings of the IEEE/CVF Conference on Computer Vision and Pattern Recognition}, 2022, pp. 5227--5237.

\bibitem{chen2022adaptformer}
Shoufa Chen, Chongjian Ge, Zhan Tong, Jiangliu Wang, Yibing Song, Jue Wang, and Ping Luo,
\newblock ``Adaptformer: Adapting vision transformers for scalable visual recognition,''
\newblock {\em Advances in Neural Information Processing Systems}, vol. 35, pp. 16664--16678, 2022.

\bibitem{kannan2019large}
Anjuli Kannan, Arindrima Datta, Tara~N Sainath, Eugene Weinstein, Bhuvana Ramabhadran, Yonghui Wu, Ankur Bapna, Zhifeng Chen, and Seungji Lee,
\newblock ``Large-scale multilingual speech recognition with a streaming end-to-end model,''
\newblock {\em arXiv preprint arXiv:1909.05330}, 2019.

\bibitem{winata2020adapt}
Genta~Indra Winata, Guangsen Wang, Caiming Xiong, and Steven Hoi,
\newblock ``Adapt-and-adjust: Overcoming the long-tail problem of multilingual speech recognition,''
\newblock {\em arXiv preprint arXiv:2012.01687}, 2020.

\bibitem{hou2021exploiting}
Wenxin Hou, Han Zhu, Yidong Wang, Jindong Wang, Tao Qin, Renjun Xu, and Takahiro Shinozaki,
\newblock ``Exploiting adapters for cross-lingual low-resource speech recognition,''
\newblock {\em IEEE/ACM Transactions on Audio, Speech, and Language Processing}, vol. 30, pp. 317--329, 2021.

\bibitem{otake2023parameter}
Shinta Otake, Rei Kawakami, and Nakamasa Inoue,
\newblock ``Parameter efficient transfer learning for various speech processing tasks,''
\newblock in {\em ICASSP 2023-2023 IEEE International Conference on Acoustics, Speech and Signal Processing (ICASSP)}. IEEE, 2023, pp. 1--5.

\bibitem{peng2023parameter}
Junyi Peng, Themos Stafylakis, Rongzhi Gu, Old{\v{r}}ich Plchot, Ladislav Mo{\v{s}}ner, Luk{\'a}{\v{s}} Burget, and Jan {\v{C}}ernock{\`y},
\newblock ``Parameter-efficient transfer learning of pre-trained transformer models for speaker verification using adapters,''
\newblock in {\em ICASSP 2023-2023 IEEE International Conference on Acoustics, Speech and Signal Processing (ICASSP)}. IEEE, 2023, pp. 1--5.

\bibitem{yang21c_interspeech}
Shu wen Yang, Po-Han Chi, Yung-Sung Chuang, Cheng-I~Jeff Lai, Kushal Lakhotia, Yist~Y. Lin, Andy~T. Liu, Jiatong Shi, Xuankai Chang, Guan-Ting Lin, Tzu-Hsien Huang, Wei-Cheng Tseng, Ko~tik Lee, Da-Rong Liu, Zili Huang, Shuyan Dong, Shang-Wen Li, Shinji Watanabe, Abdelrahman Mohamed, and Hung yi~Lee,
\newblock ``{SUPERB: Speech Processing Universal PERformance Benchmark},''
\newblock in {\em Proc. Interspeech 2021}, 2021, pp. 1194--1198.

\bibitem{nagrani2017voxceleb}
Arsha Nagrani, Joon~Son Chung, and Andrew Zisserman,
\newblock ``Voxceleb: A large-scale speaker identification dataset,''
\newblock {\em Proc. Interspeech 2017}, pp. 2616--2620, 2017.

\bibitem{sang2020open}
Mufan Sang, Wei Xia, and John~H.L. Hansen,
\newblock ``Open-set short utterance forensic speaker verification using teacher-student network with explicit inductive bias,''
\newblock {\em Proc. Interspeech 2020}, pp. 2262--2266, 2020.

\bibitem{kingma2014adam}
Diederik~P Kingma and Jimmy Ba,
\newblock ``Adam: A method for stochastic optimization,''
\newblock {\em arXiv preprint arXiv:1412.6980}, 2014.

\end{thebibliography}

% \newpage
% \onecolumngrid
% \appendix
% \section*{Supplementary Material}

\end{document}